\documentclass[twocolumn,showpacs,nofootinbib]{revtex4}

\usepackage{epsfig}
\usepackage{amsmath}

\newcommand{\req}[1]{(\ref{#1})}
\newcommand{\muR}{\mu_{R}^{2}}
\newcommand{\muF}{\mu_{F}^{2}}
\newcommand{\muO}{\mu_{0}^{2}}

\begin{document}

\title{
A note on the factorization scale dependence of the PQCD
predictions for exclusive processes 
}

\author{B. Meli\'{c}}
\thanks{Alexander von Humboldt Fellow.
On leave of absence from the 
Rudjer Bo\v{s}kovi\'{c} Institute,
Zagreb, Croatia.}
\affiliation{Institut f\"{u}r Physik,
        Universit\"{a}t Mainz,\\
        D-55099 Mainz, Germany} 
\affiliation{Institut f\"{u}r Theoretische Physik,
        Universit\"{a}t W\"{u}rzburg,\\
        D-97074 W\"{u}rzburg, Germany} 
\author{B. Ni\v{z}i\'{c}}
\author{K. Passek}
\affiliation{Theoretical Physics Division, 
        Rudjer Bo\v{s}kovi\'{c} Institute, \\
        P.O. Box 180, HR-10002 Zagreb, Croatia}
\vspace*{1.5cm}
\begin{abstract}
\vspace*{0.5cm}
We briefly review 
the calculational procedure for
the PQCD prediction for hard exclusive quantities
and reconsider the problem of the factorization scale dependence. 
\end{abstract}


\maketitle

\section{Introduction}
\label{sec:intro}
The application of perturbative QCD (PQCD) to exclusive processes  
at large momentum transfer 
is based on the factorization theorems
\cite{LeBr79etc,EfR80etc,DuM80etc,LeBr80}. 
The main idea is the separation of short- from long-distance 
effects in a sense that the high-energy region, 
being highly off-shell, is factorized from the low-energy region, 
which is characteristic of the bound-state formation. 
The factorization may be carried out order by order in perturbation theory. 
The information concerning  the long-distance dynamics 
is accumulated in the distribution amplitude (DA), 
one for each hadron involved, 
whereas the short-distance dynamics is represented 
by the hard-scattering amplitude. 
The separation of the short- from the long-distance part occurs at 
the factorization scale which is usually chosen by convenience.  
Furthermore, PQCD calculation to the finite order necessarily requires 
the renormalization of the UV divergences 
and introduces therefore the renormalization scale dependence in the 
final result.  

As is well known, one of the most critical problems in making reliable 
PQCD predictions for exclusive processes at large momentum transfer is 
how to deal with the dependence of the corresponding truncated perturbation 
series on the choice of 
the scheme for 
the QCD running coupling constant $\alpha_S(\mu_R^2)$ and on the choice  
of the renormalization scale $\mu_R$, 
as well as,  
the factorization scale $\mu_F$. 
Although the physical quantities depend neither on the
renormalization nor on the factorization scale, 
the PQCD prediction at the finite order bears the residual 
dependence on the renormalization and factorization scales, 
the choice of which introduces theoretical uncertainties in the prediction.

A lot of work has been devoted to
the analysis of the renormalization scale and scheme dependence 
\cite{FAC,PMS,BLM,BrJPR98}.
The problem of finding the optimal renormalization scale
in a given scheme has been widely discussed in the literature
and, apart from the pragmatical choice 
$\mu_R^2$ equal the characteristic scale of the process,
 three quite different approaches have been proposed:
the principle of fastest apparent convergence (FAC) \cite{FAC}, 
the principle of minimal sensitivity (PMS) \cite{PMS} 
and the Brodsky-Lepage-Mackenzie (BLM) scale setting \cite{BLM}.
A physically motivated formalism in which any two
perturbatively calculable observables can be related to each other
without any renormalization scale or scheme ambiguity has been
developed \cite{BrJPR98}. 

In contrast to the renormalization scale, 
somewhat less attention 
has been paid to the role played by the  
factorization scale.
Although one can encounter in the literature
several extensions of the treatments of the renormalization scale
to the treatments of the factorization scale
\cite{NakkagawaN82etc}, 
when examining the hard exclusive processes
a convenient choice $\mu_F^2$ equal the characteristic scale
of the process, i.e., the large momentum transfer denoted by, say 
$Q^2$ is mainly used, with the justification 
that for such a choice $\ln(Q^2/\mu_F^2)$ logarithms, 
giving rise to the growth
of the coefficients in the expansion 
of the hard-scattering amplitude when $Q^2 \gg \muF$, vanish. 
Obviously, the result will differ for some 
other choice of $\muF$. Then one can try, similarly to the 
renormalization scale problem, to justify other choices for 
the factorization scale by examining 
the underlying dynamics of the process
 \cite{DiR81,MNP99}. 

In this paper we review  the prescription and the ingredients
of the higher-order calculation of the hard exclusive
quantities 
(obtained in the, so called, standard hard-scattering approach
\cite{LeBr79etc,EfR80etc,DuM80etc,LeBr80}) 
and we reexamine their factorization scale dependence.
We show that the residual factorization scale 
dependence in the finite order of PQCD calculation reflects the 
failure of the proper resummation of all $\ln(\mu_F^2)$ logarithms. 
Thus, 
taking into account the 
factorization scale dependence of the hard-scattering amplitude 
and of the distribution amplitude by consistently
including all terms 
that are effectively of the same 
order in $\alpha_S$,  
the PQCD prediction for 
an exclusive one-scale process is free of any residual dependence 
on the factorization scale at every order of the PQCD calculation. 
The unavoidable theoretical uncertainty 
of a particular order of the PQCD calculation remains to be only due to the 
renormalization procedure. 
Nevertheless, we comment the problems one is left with when 
adopting such a procedure, especially in the case of multi-scale
processes.

In Sec. \ref{sec:sHSA} we introduce the ingredients of the
standard hard-scattering picture on the example of the
pion transition form factor, while in Sec. \ref{sec:calc}
the higher-order calculational procedure is outlined.
The discussion of the factorization scale dependence 
is given in Sec. \ref{sec:muF}. Section \ref{sec:concl}
is devoted to concluding remarks. 

\section{Standard hard-scattering picture at higher-orders}
\label{sec:sHSA}
For definiteness, notational simplicity, and clarity of
presentation, we 
consider the high-energy behavior of 
the simplest exclusive 
quantity, 
the pion transition form factor 
$F_{\gamma \pi}(Q^2)$, 
defined in terms
of the 
$\gamma^*(q,\mu) + \gamma(k,\nu) \rightarrow \pi(P)$
amplitude.
For large  momentum transfer $Q^2(=-q^2)$, 
the general factorization formula 
\cite{LeBr79etc,EfR80etc,DuM80etc,LeBr80} 
 for $F_{\gamma \pi}(Q^2)$ reads 
\begin{equation}
    F_{\gamma \pi}(Q^{2})= 
       \Phi^{*}(x,\muF) \, \otimes \, T_{H}(x,Q^{2},\muF) 
           \, .  
\label{eq:tffcf}
\end{equation}
Here, 
$\Phi(x,\muF)$ is the pion distribution amplitude; 
$T_{H}(x,Q^{2},\muF)$ is the hard-scattering amplitude; 
$\muF$ is the factorization scale, and 
$x$ denotes the pion constituent's momentum
fraction, while 
$\otimes \equiv \int_0^1 dx$. 

The hard-scattering amplitude (HSA) $T_{H}$ can be 
explicitly calculated in 
perturbation theory and represented as a
series in the QCD running coupling constant $\alpha_S(\muR)$
by
\begin{eqnarray}
  T_{H}(x,Q^2,\muF) &=&
         T_{H}^{(0)}(x,Q^2)
         + \frac{\alpha_{S}(\muR)}{4 \pi} \, 
             T_{H}^{(1)}(x, Q^2,\muF) 
             \nonumber \\ & &
         + \frac{\alpha_{S}^2(\muR)}{(4 \pi)^2} \, 
             T_{H}^{(2)}(x, Q^2,\muF,\muR) 
                + \cdots  \, , 
             \nonumber \\
\label{eq:TH}
\end{eqnarray}
where $\muR$ is the renormalization scale. 
The dependence of the 
coefficients of the expansion \req{eq:TH}
on the scales $\muR$ and $\muF$ is of the 
form 
$\ln^n(\muR/Q^2)$ and $\ln^m(\muF/Q^2)$, respectively. 

The pion distribution amplitude $\Phi(x,\muF)$, although intrinsically 
nonperturbative, satisfies the Brodsky-Lepage (BL) evolution equation 
\begin{equation}
  \muF \frac{\partial}{\partial \muF} \Phi(x,\muF)   =
   V(x,u,\muF) \, \otimes \, \Phi(u,\muF)
         \, ,
\label{eq:eveq}
\end{equation}
where $V(x,u,\muF)$ is the 
perturbatively calculable evolution kernel
\begin{eqnarray}
  \lefteqn{V(x,u,\muF)}
           \nonumber \\ &= & 
        \frac{\alpha_S(\muF)}{4 \pi} \, V_1(x,u) +
                 \frac{\alpha_S^2(\muF)}{(4 \pi)^2}  V_2(x,u) +
                 \cdots \, . \quad
\label{eq:kernel}
\end{eqnarray}
The solution of Eq. \req{eq:eveq} can be represented
as 
\begin{equation}
   \Phi(x, \muF) =
   \Phi^{LO}(x, \muF)
 + \frac{\alpha_S(\muF)}{4 \pi} \;
   \, \Phi^{NLO}(x, \muF) + \cdots
          \, ,
\label{eq:PhiLONLO}
\end{equation}
where $\Phi^{LO}$ and $\Phi^{NLO}$ denote
the leading order (LO) and next-to-leading order (NLO) 
parts, respectively.
When convoluting  the finite-order results 
\req{eq:TH} and \req{eq:PhiLONLO}
according to \req{eq:tffcf},
one is usually left with the residual dependence on
both $\muR$ and $\muF$.
The origin of the latter will be explained in the 
following.

\section{Calculational procedure}
\label{sec:calc}

In order to be able to examine the origin of the
residual dependence on $\muF$,
we first reexamine the calculational
procedure and the ingredients
of the standard hard-scattering picture
for $F_{\gamma \pi}(Q^2)$. 

The HSA $T_H$ 
is obtained by evaluating 
the $\gamma^* + \gamma \rightarrow q \overline{q}$
amplitude, which we denote by $T$. 
Owing to the fact that final-state quarks are taken to be
massless and on-shell, the amplitude 
contains collinear singularities.
Since $T_H$ is a finite quantity by definition,
collinear singularities have to be subtracted.
Therefore, $T$ factorizes as 
\begin{equation}
    T(u,Q^2) = T_H(x, Q^2, \muF) \, \otimes \, Z_{T,col}(x, u; \muF)
        \, ,
\label{eq:TTHZ}
\end{equation}
with collinear singularities being subtracted at the scale 
$\muF$ and absorbed into the constant $Z_{T,col}$. 
The UV singularities are removed by
the renormalization of the fields
and by the coupling-constant renormalization at the 
(renormalization) scale $\muR$. 

The process-independent pion DA 
in a frame where
$P^+=P^0+P^3=1$, $P^-=P^0-P^3=0$, and $P_{\perp}=0$
is defined \cite{LeBr80,Ka85etc,BrD86} 
as
\begin{eqnarray}
 \Phi (u)
 & = &  \int \frac{dz^-}{2 \pi} e^{i(u-(1-u))z^- /2}
      \nonumber \\ & & \times
    \left< 0 \left| 
  \bar{\Psi}(-z) \, \frac{\gamma^+ \gamma_5}{2\sqrt{2}}
           \, \Omega \, \Psi(z) 
    \right| \pi \right> _{(z^+=z_{\perp}=0)}
                , 
\label{eq:PhiOPi}
\end{eqnarray}
where 
$ \Omega  =  
  \mbox{exp} \left\{ i g \int_{-1}^{1} ds A^+(z s)z^-/2 \right\} $
is a path-ordered factor
making $\Phi$ gauge invariant. 
Owing to the light-cone singularity at $z^2=0$ \cite{LeBr80,BrD86} 
the matrix element in \req{eq:PhiOPi} is UV divergent. 
After regularization and renormalization at the scale $\tilde{\mu}_R^2$, 
$z^2$ is effectively smeared over a region of order
$z^2=-z_{\perp}^2\sim 1/\tilde{\mu}_R^2$.
As a result, a finite quantity, namely, 
the pion DA $\Phi(v,\tilde{\mu}_R^2)$, 
is obtained and corresponds to 
the pion wave function integrated
over the pion intrinsic transverse momentum up to the scale 
$\tilde{\mu}_R^2$.

The pion DA as given in
\req{eq:PhiOPi}, with $\left| \pi \right>$ being the physical
pion state, cannot be determined by perturbation theory. 
If the meson state $\left| \pi \right>$ is replaced
by a $\left| q \overline{q}; t \right>$ state composed of a 
free (collinear, massless, and on-shell) quark and antiquark 
(carrying momenta $t P$ and $(1-t) P$ and pseudoscalar meson
quantum numbers),
then the amplitude \req{eq:PhiOPi} becomes
\begin{eqnarray}
  \tilde{\phi} (u, t)
& = &  \int \frac{dz^-}{2 \pi} e^{i(u-(1-u))z^- /2}
   \nonumber \\ & & \times
    \left< 0 \left| 
  \overline{\Psi}(-z) \, \frac{\gamma^+ \gamma_5}{2\sqrt{2}}
           \, \Omega \, \Psi(z) \right| q \overline{q}; t \right>
              \, . \quad
\label{eq:phiOqq}
\end{eqnarray}
Taking \req{eq:phiOqq} into account,
we can express Eq. \req{eq:PhiOPi} as 
\begin{equation}
   \Phi(u)  = 
       \tilde{\phi}(u,t) \otimes
            \, \left< q\bar{q}; t | \pi \right> 
               \, .
\label{eq:Phitphirest}
\end{equation}
The distribution $\tilde{\phi}(u,t)$ can be treated perturbatively,
which enables us to investigate
the high-energy tail of
the pion DA and its evolution.
The $\tilde{\phi}(u,t)$ distribution is
multiplicatively renormalizable 
owing to the multiplicative renormalizability
of the composite operator
  $\overline{\Psi}(-z) \, \gamma^+ \gamma_5
           \, \Omega \, \Psi(z)$. 
This means that the UV singularities 
that are not removed by the renormalization of the
fields and by  
the coupling-constant renormalization 
factorize in the renormalization constant 
$Z_{\phi,ren}$
at the (renormalization) scale $\tilde{\mu}_R^2$. 
Apart from UV singularities, the matrix element in 
\req{eq:phiOqq} contains also collinear singularities. 
Subtracting these singularities at the scale $\mu_0^2$ and 
absorbing in $Z_{\phi,col}$, 
we can write Eq. \req{eq:phiOqq}
as 
\begin{eqnarray}
   \lefteqn{\tilde{\phi}(u,t)} \nonumber \\
& = & Z_{\phi,ren}(u,v; \tilde{\mu}_R^2) \otimes 
              \phi_V(v,s; \tilde{\mu}_R^2, \mu_0^2) \otimes
               Z_{\phi,col}(s,t; \mu_0^2)
                 \, .
           \nonumber \\ 
\label{eq:ZfVZ}
\end{eqnarray}

By combining \req{eq:Phitphirest} and \req{eq:ZfVZ},
we obtain the distribution $\Phi(u)$ in the form
\begin{eqnarray}
   \Phi(u) & = &
            Z_{\phi,ren}(u,v; \tilde{\mu}_R^2) \otimes
              \Phi(v, \tilde{\mu}_R^2)
             \, ,
\label{eq:PhiZPhi}
\end{eqnarray}
where 
\begin{equation}
   \Phi(v,\tilde{\mu}_R^2)=
         \phi_V(v,s; \tilde{\mu}_R^2, \mu_0^2) \otimes
         \Phi(s, \mu_0^2)
           \, .
\label{eq:PhifVPhi}
\end{equation}
Here, 
\begin{equation}
\Phi(s, \mu_0^2) = Z_{\phi,col}(s,t; \mu_0^2) \otimes 
 \, \left< q\bar{q}; t | \pi \right> 
               \, 
\label{eq:PhiNP}
\end{equation}
represents the nonperturbative input (containing 
collinear singularities and all effects of confinement and pion bound-state 
dynamics) determined at the scale 
$\mu_0^2$, while 
$\phi_V(v,s; \tilde{\mu}_R^2, \mu_0^2)$
governs the evolution of $\Phi(v,\mu_0^2)$ 
to the scale $\tilde{\mu}_R^2$.
By differentiating \req{eq:PhiZPhi}
with respect to $\tilde{\mu}_R^2$, one obtains
\req{eq:eveq},
with the evolution potential $V$ given by%
\begin{equation}
   V(\tilde{\mu}_R^2) = -Z_{\phi, ren}^{-1}(\tilde{\mu}_R^2) 
          \, \left( \tilde{\mu}_R^2
       \frac{\partial}{\partial \tilde{\mu}_R^2} Z_{\phi, ren}(\tilde{\mu}_R^2)
              \right)
          \, .
\label{eq:VZ}
\end{equation}
To simplify the expressions, 
the convolution ($\otimes$) is here, and where appropriate
replaced by the matrix multiplication in $x$-$y$ space 
(the unit matrix is defined as $\openone = \delta(x-y)$),
while the $x$, $y$ variables are suppressed. 

By convoluting 
the amplitudes $T(u,Q^2)$ and $\Phi(u)$, 
Eqs. \req{eq:TTHZ} and \req{eq:PhiZPhi}, respectively, 
in analogy with \cite{CuF80,LeBr80} we obtain 
the pion transition form factor $F_{\gamma \pi}(Q^2)$: 
\begin{equation}
  F_{\gamma \pi}(Q^2) =
       \Phi^{\dagger}(u) \, \otimes \,  T(u, Q^2)
          \, .
\label{eq:Fpiur1}
\end{equation}
Now, in order that the factorization
holds, $\tilde{\mu}_R^2$ has to to coincide with $\muF$ 
and by making use of the fact that%
\begin{equation}
      Z_{T,col}(x,u; \muF) \otimes Z_{\phi, ren}(u,v; \muF)
           = \delta(x-v)
              \, ,
\label{eq:ZTZf}
\end{equation}
the divergences of $T(u,Q^2)$ and $\Phi(u)$ in \req{eq:Fpiur1} cancel 
(this has been explicitly shown in \cite{MNP01} 
up to $n_f$ proportional terms of $O(\alpha_s^2)$)
and we are left with
the finite perturbative expression  for the pion
transition form factor \req{eq:tffcf}.

We note here that the same
factorization (and renormalization) scheme
is employed in the hard-scattering and DA part, i.e.,
in Eqs. \req{eq:TTHZ} and \req{eq:PhiZPhi}, respectively.
Furthermore, as pointed out in \cite{MelicMP03},
the evolution equation as defined by
\req{eq:eveq} and \req{eq:kernel}
corresponds to the simplified scheme fixed by the
preference that the distribution amplitude should have no 
dependence on the renormalization scale\footnote{Note that, in general,
such a residual dependence appears along with 
the evolution kernel depending on two scales:
\begin{eqnarray*}
\lefteqn{V(x,u, \muF)} \\
 &= &\frac{\alpha_s(\muR)}{4\pi} V_1(x,u) 
+ \frac{\alpha_s^2(\muR)}{(4\pi)^2}  
\left( V_2(x,u) 
- \beta_0 V_1 (x,u) \ln
  \left(\frac{\muR}{\muF}\right) \right) 
\\ & &
+ \, O\left(\alpha_s^3\right) \, .
\end{eqnarray*}
Here $\muR$ corresponds to the scale of the coupling constant, while $\tilde{\mu}_R^2 = \muF$
denotes the scale at which remaining UV divergences, 
due to the renormalization of the composite operator,factorize.
}. 

It is worth pointing out that
the scale $\muF$ representing the boundary between the low- and 
high-energy parts in
\req{eq:tffcf} plays the role of the separation scale for
collinear singularities
in $T(u,Q^2)$, on one hand,
and of the renormalization scale for
UV singularities appearing in the perturbatively calculable part
of the distribution amplitude $\Phi(u)$, on the other hand.

\begin{figure}
  \centerline{\epsfig{file=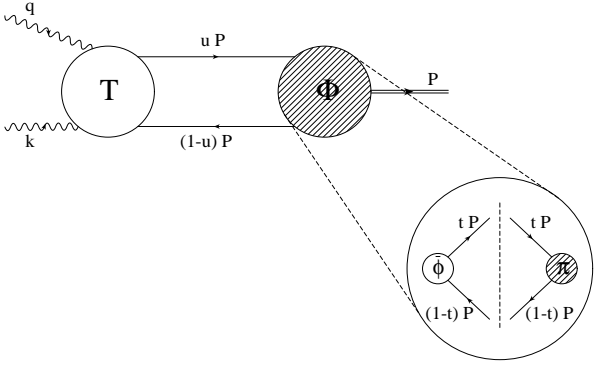,height=150pt,width=240pt,silent=}}
 \caption{Pictorial representation of the 
    calculational ingredients
    of the pion transition form factor. 
    $T$ represents the perturbatively calculable 
    $\gamma^* + \gamma \rightarrow q \overline{q}$
    amplitude , while $\Phi$ denotes the (unrenormalized)
    pion distribution
    amplitude given by \protect\req{eq:PhiOPi}, which can be
    expressed, as in \protect\req{eq:Phitphirest},
    in terms of the perturbatively calculable part
    $\tilde{\phi}$
    \protect\req{eq:phiOqq}
    and the perturbatively incalculable part.}  
 \label{f:FpiDA}
\end{figure}

The calculational procedure explained above is 
illustrated in Fig. \ref{f:FpiDA}.
%

\section{Factorization scale dependence}
\label{sec:muF}

We next turn to the discussion of the $\mu_F^2$ dependence of the 
pion transition form factor defined as in \req{eq:tffcf}. 

Concerning the pion distribution amplitude $\Phi(x,\muF)$, its 
dependence on $\mu_F^2$ is specified 
by the evolution equation \req{eq:eveq} and,
as can be seen from Eq. \req{eq:PhifVPhi},
this dependence is completely contained in the evolutional
part $\phi_V$.  
By calculating the perturbatively obtainable amplitude 
$\tilde{\phi}$ \cite{MNP01} directly from \req{eq:phiOqq},
the result obtained for $\phi_V$ can be organized as 
\begin{eqnarray}
\phi_V(\muF,\muO) & = &
      \phi_{V}^{LO}(\muF,\muO)
 + \frac{\alpha_S(\muF)}{4 \pi}
      \phi_{V}^{NLO}(\muF,\muO) 
         \nonumber \\ & & + \cdots
             \, ,
\label{eq:phiVevLONLO}
\end{eqnarray}
where
\begin{subequations}
\begin{eqnarray}
 \phi_{V}^{LO}(\muF,\muO) &=&
     \openone
 + \frac{\alpha_S(\muF)}{4 \pi}
    \ln \frac{\muF}{\mu_0^2} \, V_1
       \nonumber  \\ & &
 + \frac{\alpha_S^2(\muF)}{(4 \pi)^2}
    \ln^2 \frac{\muF}{\mu_0^2}\,
   \frac{1}{2} \, (V_1^2 + \beta_0 \, V_1)
   \nonumber \\ & & + \cdots \, ,
    \label{eq:phiVevLO} \\
\phi_{V}^{NLO}(\muF,\muO) &=&
  \frac{\alpha_S(\muF)}{4 \pi}
    \ln \frac{\muF}{\mu_0^2}
           \, V_2 \,
      + \cdots
           \, ,
    \label{eq:phiVevNLO}
\end{eqnarray}
\label{eq:phiVev}
\end{subequations}
and functions $V_n$ represent the $n$-loop evolutional kernels
appearing in Eq. \req{eq:eveq}. 
The terms explicitly given in \req{eq:phiVev}
correspond to the results of the two-loop calculation \cite{MNP01}. 
In writing \req{eq:phiVev}, 
the use has been made of
\begin{displaymath}
\frac{\alpha_S(\muF)}{4 \pi} 
  \ln \frac{\muF}{\mu_0^2} 
  \approx \frac{1}{\beta_0} \left( 1 - 
\frac{\alpha_S(\muF)}{\alpha_S(\mu_0^2)} \right)
 = O(\alpha_S^0)
     \, .
\end{displaymath}

On the other hand,
the complete LO and NLO behavior of $\phi_V(v,s; \muF, \muO)$
and, consequently, of $\Phi(v,\muF)$
can be determined by solving the
evolution equation \req{eq:eveq} or equivalently
\begin{eqnarray}
       \lefteqn{\muF 
       \frac{\partial}{\partial \muF} \phi_V (v,s,\muF,\muO)}
           \nonumber \\  
        \qquad & \qquad =& V(v,s',\muF) \, \otimes \, \phi_V(s',s,\muF,\muO)
             \, . \qquad 
\label{eq:VfV}
\end{eqnarray}
The LO result is of the form 
\begin{eqnarray}
 \phi_V^{LO}(v,s; \muF, \muO) &=&
      \sum_{n=0}^{\infty} {}' 
         \,
      \frac{v (1-v)}{N_n}
         \,
        C_n^{3/2}(2 v-1)  
    \nonumber \\ &  & \times \, 
        C_n^{3/2}(2 s -1) \;
        \left(
    \frac{\alpha_S(\muF)}{\alpha_S(\mu_0^2)}
        \right)^{-\gamma_n^{(0)}/\beta_0}
          \hspace{-1.0cm} , \hspace{1.0cm}
\label{eq:phiVLOcom}
\end{eqnarray}
where $N_n=(n+1)(n+2)/(4 (2 n+3) )$,
while $C_n^{3/2}(2 x -1)$ are the Gegenbauer polynomials (
the eigenfunctions of the LO kernel $V_1$
with the corresponding eigenvalues $\gamma_n^{(0)}$ \cite{MNP99}).
The above given complete LO prediction 
represents the summation of all
$(\alpha_S \ln \muF/\mu_0^2)^n$ terms from \req{eq:phiVevLO}.
The complete formal solution of the NLO evolution equation
was obtained in \cite{Mu94etc} by using conformal constraints
and the form of $\phi_V^{NLO}$ (corresponding
to the resummation of \req{eq:phiVevNLO}) can
be extracted from the results listed in \cite{MNP99}.

It is important to realize that
the method employed above to study the $\mu_F^2$ behavior of 
$\phi_V$ can be used to examine 
the dependence of the hard-scattering amplitude $T_{H}(x,Q^{2},\muF)$
on the scale $\muF$, as well.  

By differentiating \req{eq:tffcf} with respect to $\muF$ and
by taking into account \req{eq:eveq},
one finds that the hard-scattering amplitude satisfies
the evolution equation
\begin{equation}
  \muF \frac{\partial}{\partial \muF} 
        T_H(x, Q^2, \muF) = -
        T_H(y, Q^2, \muF)  \, \otimes \,
            V(y,x;\muF) 
         \, . 
\label{eq:EvEqT}
\end{equation}
This equation%
\footnote{The Eq. \req{eq:EvEqT} can be also obtained by combining 
Eq. \req{eq:TTHZ} with Eqs. \req{eq:VZ} and 
\req{eq:ZTZf}.}
is analogous to the DA evolution equation 
\req{eq:eveq}.
Similarly to the above discussed
solution of the DA evolution equation, 
the finite-order solution of \req{eq:EvEqT}
would contain the complete dependence on $\muF$,
to given order in $\alpha_S$,
in contrast to
the expansion  \req{eq:TH}
truncated at the same order and containing unresummed logs.
Let us note that 
the explicit expressions for the hard-scattering amplitude
$T_{H}(x,Q^{2},\muF)$ in a form \req{eq:TH}, evaluated
up to $n_f$-proportional NNLO terms, are given in \cite{MNP01}.

The $\muF$ dependence of $T_H(x, Q^2, \muF)$
can be, similarly to \req{eq:PhifVPhi}, 
factorized in the function $\phi_V(y,x,Q^2,\muF)$ as
\begin{equation}
  T_H(x,Q^2,\muF) =  T_H(y,Q^2,\muF=Q^2) \otimes
                   \phi_V(y,x,Q^2,\muF)
            \, .
\label{eq:tTHfV}
\end{equation}
Using \req{eq:VfV} 
one can show by partial integration that
\req{eq:tTHfV} 
indeed represents the solution of
the evolution equation \req{eq:EvEqT}%
.

When calculating to finite order in $\alpha_S$,
it seems not quite consistent to adopt in the literature
often encountered procedure in which the
$\Phi(x,\muF)$ distribution obtained by solving 
the evolution equation \req{eq:eveq} 
is convoluted
with $T_H(x,Q^2,\muF)$ obtained by the
truncation of the expansion \req{eq:TH}. 
In the latter case, only the partial dependence
on $\muF$ is included 
(logs are not resummed),
in contrast to the former case and 
hence the residual dependence on the factorization scale $\muF$ enters. 

The proper procedure would be to convolute 
$\Phi$ \req{eq:PhifVPhi} and $T_H$  \req{eq:tTHfV} 
in terms of the same function
$\phi_V$, where $\phi_V$ can be 
given by \req{eq:phiVev} with unressummed logs
or can represent the solution of
\req{eq:VfV}, i.e., the resummed result.
In both cases the $\muF$ dependence of $\Phi$ and
$T_H$ completely cancels out and there is no residual dependence
on $\muF$.
One usually uses the resummed form of $\phi_V$ in DA $\Phi$, i.e, 
$\Phi$ is taken as a solution of evolution equation, and therefore this
procedure should be applied for $T_H$ as well.

Substituting \req{eq:PhifVPhi} and \req{eq:tTHfV} 
in \req{eq:tffcf}, we obtain
\begin{eqnarray}
  \lefteqn{F_{\gamma \pi}(Q^2)} \nonumber \\
    & = & 
   T_H(y, Q^2, Q^2) 
          \, \otimes \phi_V(y,s, Q^2, \mu_0^2) 
          \, \otimes \,\Phi^*(s, \mu_0^2)
                 \, , \qquad 
\label{eq:Fpi2q2}
\end{eqnarray}
where
\begin{equation}
   \phi_V(y,x, Q^2, \muF)\, \otimes \phi_V(x,s,\muF,\mu_0^2) 
          = \phi_V(y,s, Q^2, \mu_0^2) 
               \, 
\label{eq:fvfvfv}
\end{equation}
has been taken into account. 
It is important to realize that the expression \req{eq:fvfvfv} is valid at 
every order of a PQCD calculation%
\footnote{The Eq. \req{eq:fvfvfv} can be easily checked to the NLO order 
\cite{MNP01} 
by using  the LO result \req{eq:phiVLOcom} and the NLO results of 
Ref. \cite{Mu94etc}.}, and hence even the finite order prediction
for $F_{\gamma \pi}(Q^2)$ does not depend on the choice of the 
$\muF$ scale%
\footnote{
Let us, following Eqs. \req{eq:kernel} and \req{eq:phiVevLONLO},
define the finite  order quantities
\begin{eqnarray*}
\lefteqn{\phi_V^{(n)}(\muF,\muO)} \nonumber \\
       & = &
      \phi_{V}^{LO}(\muF,\muO)
 + \cdots + \frac{\alpha_S^{n}(\muF)}{(4 \pi)^{n}}
      \phi_{V}^{N \cdots NLO}(\muF,\muO) 
             \, , \quad
       \nonumber \\ & &
\label{eq:phiVevLONLOfin}
\end{eqnarray*}
and
\begin{displaymath}
  V^{(n)}(\muF) = 
        \frac{\alpha_S(\muF)}{4 \pi} \, V_1 +
                    \cdots
               +  \frac{\alpha_S^{n+1}(\muF)}{(4 \pi)^{n+1}}  V_n 
                  \,  \quad
\label{eq:kernelfin}
\end{displaymath}
(here $n=0, \ldots $).
The functions $\phi_V^{(n)}(Q^2, \muF)$ and $\phi_V^{(n)}(\muF,\mu_0^2)$
represent the solutions of the evolutional
equations
\begin{eqnarray*}
 \muF \frac{\partial}{\partial \muF} \phi_V^{(n)} (\muF,\muO)
        & = & V^{(n)}(\muF) \, \otimes \, \phi_V^{(n)}(\muF,\muO)
             \, , \nonumber \\
             & & \\
 \muF \frac{\partial}{\partial \muF} \phi_V^{(n)} (Q^2,\muF)
        & = & - \phi_V^{(n)}(Q^2,\muF)\, \otimes \, V^{(n)}(\muF) 
             \, . \qquad 
             \nonumber \\
\end{eqnarray*}
It is now easy to prove that the convolution
\begin{displaymath}
   \phi_V^{(n)}(Q^2, \muF)\, \otimes \phi_V^{(n)}(\muF,\mu_0^2) 
        \, 
\label{eq:fvfv}
\end{displaymath}
indeed does not depend on $\muF$.
}
.

Hence, the  expression \req{eq:fvfvfv} represents the resummation of the 
$\ln(Q^2/\mu_0^2)$ logarithms over the intermediate $\muF$ scale, 
performed in such a way that both 
the logarithms $\ln(\muF/\mu_0^2)$ originating from the 
perturbative part of the DA and the 
$\ln(Q^2/\muF)$ logarithms from the hard-scattering part are resummed. 
The effect in the final prediction,
at every order, 
is the same as if we had performed 
the complete renormalization-group resummation of $\ln(Q^2/\mu_0^2)$ 
logarithms. 

Although by using the explicit results
for \req{eq:TH} and the evolution equation solution for $\phi_V$,
it is straightforward to employ \req{eq:tTHfV}
and obtain $T_H$ with resummed $\ln(\muF/Q^2)$ logs,
it is much easier that the complete resummation is performed
in the distribution amplitude.
Hence, by adopting the common choice $\mu_F^2 = Q^2$,
we avoid the need
for the resummation of the $\ln(Q^2/\muF)$ logarithms in 
the hard-scattering part, making the calculation simpler 
and hence, for practical purposes, the preferable form of
$F_{\gamma \pi}(Q^{2})$ is given by
\begin{equation}
    F_{\gamma \pi}(Q^{2})= 
   T_H(x, Q^2, Q^2) 
        \, \otimes \,  \Phi^{*}(x,Q^2)
                    \,. 
\label{eq:tffcfN}
\end{equation}
We stress here that in this approach, in which the consistent treatment
of $\Phi$ and $T_H$ dependence on $\muF$ is required,
any other choice of $\muF$ would lead to the same result,
only the calculation would be more involved.

\section{Concluding remarks}
\label{sec:concl}

We have sketched the higher-order PQCD calculational
procedure
for the hard exclusive quantities on the example of
the pion transition form factor $F_{\gamma \pi}$.

Furthermore, we have argued that the $F_{\gamma \pi}$
prediction \req{eq:tffcf}
is independent 
of the factorization scale $\muF$  
at every order in $\alpha_S$,
when  both the hard scattering part $T_H$ and the distribution 
amplitude $\Phi$ are 
consistently treated regarding the $\muF$ dependence,
i.e., in both quantities the $\ln \muF$ logarithms are 
resummed or in both quantities they are not resummed. 
The $\muF$ dependence of $\Phi$ then exactly cancels the
$\muF$ dependence of $T_H$, the
choice of the factorization scale is therefore nonessential 
and the predictions obtained by using any choice of $\muF$ 
are equal to the results obtained using, 
for practical purposes the simplest intermediate choice $\muF=Q^2$,
where $Q^2$ represents the characteristic scale of the process. 

The true expansion parameter left is 
$\alpha_S(\muR)$, 
with $\muR$ representing
the renormalization scale of the complete
perturbatively calculable part of the pion
transition form factor \req{eq:Fpi2q2}, i.e., 
of
\begin{equation} 
T_H(s, Q^2,\muO)=
 T_H(y, Q^2, Q^2) \otimes \phi_V(y,s, Q^2, \muO)
    \, .
\label{eq:pertpart}
\end{equation} 
Therefore, although 
$F_{\gamma \pi}(Q^2)$
depends exclusively on the characteristic scale of the process $Q^2$,
we are left with
the residual dependence on the $\muR$ scale, 
when calculating to finite order.
The intermediate 
scale at which the short- and long-distance dynamics separate, 
the factorization scale  
disappears from the final prediction at every order in $\alpha_S$
and therefore does not introduce any theoretical uncertainty into the 
PQCD calculation for exclusive processes. 

Above discussed calculational prescription for the factorization scale 
independent calculation is also upheld for other PQCD exclusive 
one-scale processes. 
However, in the case of exclusive processes which involve 
more than one typical scale the treatment of the factorization scale 
dependence is more involved.
The subtlety in the preceding approach lies in the fact that we have
traded one dependence on $\mu_F$ for another one.
Namely, the choice we have made there is that we resum
$\ln$ terms up to the relevant scale of the process $Q^2$.
Although, this might seem reasonable for the one external scale
processes, 
such as the one which define
$F_{\gamma \pi}$ ($\gamma \gamma^* \to \pi$)
or the pion electromagnetic form factor, in processes
with two scales, for example in 
the one in which the 
general pion transition form factor 
$F_{\gamma^* \pi}$ ($\gamma^* \gamma^* \to \pi$) appears,
one immediately encounters the ambiguity 
of how to choose the relevant scale
up to which the logs will be resummed. 
The existence of such ambiguities 
seems to be unavoidable artefact of the PQCD
calculation.

\begin{acknowledgments}
  We would like to thank P. Kroll, D. M\"{u}ller and A. Radyushkin
  for usefull comments.
  One of us (B.M.) acknowledges the support 
  by the Alexander von Humboldt Foundation and the hospitality
  of the theory groups at the Institut f\"{u}r Physik,
  Universit\"{a}t Mainz \& Institut f\"{u}r Theoretische Physik,
  Universit\"{a}t W\"{u}rzburg.
  This work was supported by the Ministry of Science and Technology
  of the Republic of Croatia under Contract No. 00980102.
\end{acknowledgments}


\newpage

\begin{thebibliography}{10}

\bibitem{LeBr79etc}
G.~P.~Lepage and S.~J.~Brodsky,
Phys.\ Lett.\ B {\bf 87}, 359 (1979),
%
Phys.\ Rev.\ Lett.\ {\bf 43}, 545 (1979).
%
\bibitem{EfR80etc}
A.~V.~Efremov and A.~V.~Radyushkin,
Theor.\ Math.\ Phys.\ {\bf 42}, 97 (1980),
%
Phys.\ Lett.\ B {\bf 94}, 245 (1980).
%
\bibitem{DuM80etc}
A.~Duncan and A.~H.~Mueller,
Phys.\ Lett.\ B {\bf 90}, 159 (1980),
%
Phys.\ Rev.\ D {\bf 21}, 1636 (1980).
%
\bibitem{LeBr80}
G.~P.~Lepage and S.~J.~Brodsky,
Phys.\ Rev.\ D {\bf 22}, 2157 (1980).
%
\bibitem{FAC}
G.~Grunberg,
%
Phys.\ Rev.\ D {\bf 29}, 2315 (1984);
%
\bibitem{PMS}
P.~M.~Stevenson,
%
%
%
Nucl.\ Phys.\ B {\bf 231}, 65 (1984).
%
\bibitem{BLM}
S.~J.~Brodsky, G.~P.~Lepage and P.~B.~Mackenzie,
Phys.\ Rev.\ D {\bf 28}, 228 (1983);
%
\bibitem{BrJPR98}
S.~J. ~Brodsky, C.-R. ~Ji, A. ~Pang and
D. ~G. ~Robertson,
Phys.\ Rev.\ D {\bf 57}, 245 (1998).
%
\bibitem{NakkagawaN82etc}
H.~Nakkagawa and A.~Niegawa,
Phys.\ Lett.\ B {\bf 119} (1982) 415;
P.~M.~Stevenson and H.~D.~Politzer,
Nucl.\ Phys.\ B {\bf 277}, 758 (1986);
C.~J.~Maxwell,
%
Nucl.\ Phys.\ B {\bf 577}, 209 (2000).
%
\bibitem{DiR81}
F.~M.~Dittes and A.~V.~Radyushkin,
Yad.\ Fiz. \ {\bf 34}, 529 (1981) 
[Sov.\ J.\ Nucl.\ Phys.\ {\bf 34}, 293 (1981)].
%
\bibitem{MNP99}
B.~Meli\'{c}, B.~Ni\v{z}i\'{c} and K.~Passek,
Phys.\ Rev.\ D {\bf 60}, 074004 (1999).
%
\bibitem{Ka85etc}
G.~R.~Katz,
Phys.\ Rev.\ D {\bf 31}, 652 (1985),
%
PhD Thesis (1986).
%
\bibitem{BrD86}
S.~J.~Brodsky, P.~Damgaard, Y.~Frishman and G.~P.~Lepage,
Phys.\ Rev.\ D {\bf 33}, 1881 (1986).
%
\bibitem{CuF80}
G.~Curci, W.~Furmanski and R.~Petronzio,
Nucl.\ Phys.\ B {\bf 175}, 27 (1980).
%
\bibitem{MNP01}
B. Meli\'{c}, B. Ni\v{z}i\'{c}, and K. Passek, 
Phys.\ Rev.\ D {\bf 65}, 053020 (2001).
%
\bibitem{MelicMP03}
B. Meli\'{c}, K. Passek-Kumeri\v{c}ki, D. M\"{u}ller, 
Phys.\ Rev.\ D {\bf 68}, 014013 (2003).
%
\bibitem{Mu94etc}
D. M\"{u}ller, 
Phys.\ Rev.\ D {\bf 49}, 2525 (1994),
%
Phys.\ Rev.\ D {\bf 51}, 3855 (1995).

\end{thebibliography}
\end{document}